# PQC: Extended Triple Decomposition Problem (XTDP) Applied To *GL(d, F$_p$)* – An Evolved Framework For Canonical Non-Commutative Cryptography

P. Hecht[1]

*Abstract*— Post-Quantum Cryptography (PQC) attempts to find cryptographic protocols resistant to attacks using Shor's polynomial time algorithm for numerical field problems or Grover's search algorithm. A mostly overlooked but valuable line of solutions is provided by non-commutative algebraic structures, specifically canonical protocols that rely on one-way trapdoor functions (OWTF). Here we develop an evolved algebraic framework who could be applied to different asymmetric protocols. The (canonic) trapdoor one-way function here selected is a fortified version of the Triple decomposition Problem (TDP) developed by Kurt. The original protocol relies on two linear and one quadratic algebraic public equation. As quadratic equations are much more difficult to cryptanalyze, an Algebraic Span Attack (ASA) developed by Boaz Tsaban, focus on the linear ones. This seems to break our previous work. As countermeasure, we present here an Extended TDP (cited as XTDP in this work). The main point is that the original public linear equations are transformed into quadratic ones and the same is accomplished for exchanged tokens between the entities. All details not presented here, could be found at the cited references.

*Keywords* – Post-Quantum Cryptography, Non-Commutative Cryptography, General Linear Group, Linear Algebra, Triple Decomposition Problem, OWTF, IND-CCA2.

## 1. INTRODUCTION.

**P**ost-Quantum Cryptography (PQC) is a trend that has an official NIST status [1,2] and which aims to be resistant to quantum computers attacks like Shor [3] and Grover [4] algorithms. NIST initiated last year a process to solicit, evaluate, and standardize one or more quantum-resistant public-key cryptographic algorithms [1]. Particularly Shor algorithm provided a quantum way to break asymmetric protocols.

Security of a canonical non-commutative protocol always relies his security on a one-way trapdoor function (OWTF) [5]. For instance, in an algebraic context, the conjugacy search problem, decomposition problem, double coset problem, triple decomposition problem, factorization search problem, commutator based or simultaneous conjugacy search problem. All are hard problems assumed to belong to AWPP time-complexity (but out of BQP), which yield convenient computational security against current quantum attacks.

In a previous work [6], we developed a framework based on the GL(d, F$_p$), working with d-dimensional non-singular matrices of elements in Z$_p$ and presented a full solution of a KEM based on p=251 field and using Kurt's TDP [5, 7]. More details could be consulted on that paper.

## 2. ALGEBRAIC SPAN ATTACK (ASA)

At time of publishing [6], we were not aware of the ASA attack developed by Tsaban and others [8]. It seems a fruitful way of cryptanalyzing non-commutative canonical protocols. As that paper declares, TDP is the only canonical protocol that is non-affected by earlier methods and that was also our own idea about it and the motivation to adopt it.

ASA concentrates on a weak point of TDP. This protocol manages public keys in format of two linear equations with two unknowns each and a quadratic one with three unknowns. Tsaban paper exposes the strong point of ASA: quadratic equations may be very difficult to solve, so he targets the linear equations with algebraic spans.

An obvious defense could be to extend TDP so that all public components are quadratic and that is the main point and purpose to present XTDP, an extension of the original framework. Further study will support the ASA resistance conjecture of XTDP or find an alternative way of cryptanalyzing it, perhaps by a new algebraic attack.

## 3. TDP REVISITED.

All details could be found at [5, 6, 7], but a short overview will help as an introduction to XTDP.

The public platform is a monoid M with two separate sets of five subsets each affected by invertibility and commutativity restrictions. In our framework, all subsets belongs to GL(d, F$_p$), the general linear group. The two sets are defined as A and B and their subsets are:

$$A : \{A_1, A_2, A_3, X_1, X_2\}$$
$$B : \{B_1, B_2, B_3, Y_1, Y_2\}$$

Subjected to following restrictions:

**(invertibility)** $\{X_1, X_2, Y_1, Y_2\}$ should be invertible. In our specific platform, all subsets are invertible.

**(commutativity)** $[A_2,Y_1]=[A_3,Y_2]=[B_1,X_1]=[B_2,X_2]=I$ in terms of group commutators.

Alice and Bob agree on using respectively A and B. All lowercase variables are random non-biased selections of uppercase sets (i.e., $a_i \in A_i$).

---
[1] Pedro Hecht: Maestría en Seguridad Informática, FCE-FCEyN-FI (Universidad de Bs Aires) phecht@dc.uba.ar

The protocol goes as follows:

(1) Alice chooses $\{a_1, a_2, a_3, x_1, x_2\}$ and computes:
$$u=a_1x_1, \quad v=x_1^{-1}a_2x_2, \quad w=x_2^{-1}a_3$$
The private key is $(a_1, a_2, a_3)$ and the public key is $(u, v, w)$.

(2) Bob chooses $\{b_1, b_2, b_3, y_1, y_2\}$ and computes:
$$p=b_1y_1, \quad q=y_1^{-1}b_2y_2, \quad r=y_2^{-1}b_3$$
The private key is $(b_1, b_2, b_3)$ and the public key is $(p, q, r)$.

(3) Both exchange public keys and compute the common key:
(Alice) $K_A = a_1 p a_2 q a_3 r = a_1 b_1 a_2 b_2 a_3 b_3$
(Bob) $K_B = u b_1 v b_2 w b_3 = a_1 b_1 a_2 b_2 a_3 b_3$

As could be verified, $\{u, w, p, r\}$ are lineal equations and $\{v, q\}$ are quadratic. ASA avoid attacking the non-linear equations and target the linear ones.

## 4. XTDP.

This protocol transforms TDP into a stronger one changing all linear public equations into quadratic ones in a mostly symmetric way. This feature enforces true TDP solving of each public piece to cryptanalyze it. As a drawback, in case of volatile or session public keys, it needs a double pass of public information. Aside from this not too disturbing disadvantage, the XTDP based framework could be a fast and secure way to obtain a canonical non-commutative PQC solution.

The public platform is the general linear group $GL(d, F_p)$ with two separate sets of seven disjoint subsets each affected by commutativity restrictions, as invertibility is a-priori assured. In our framework, the two sets are defined as A and B and their subsets are:

$$A : \{A_1, A_2, A_3, X_0, X_1, X_2, X_3\}$$
$$B : \{B_1, B_2, B_3, Y_0, Y_1, Y_2, Y_3\}$$

Subjected to following restrictions:

**(commutativity)**
$[A_1,Y_0]=[A_2,Y_1]=[A_3,Y_2]=[B_1,X_1]=[B_2,X_2]=[B_3,X_3]=I$ in terms of group commutators.

Alice and Bob agree on using respectively A and B. All lowercase variables are random non-biased selections of uppercase sets (i.e., $a_i \in A_i$). It is crucial to obtain IND-CCA2 semantic security that all instances are randomly selected.

(4) Alice chooses $\{a_1, a_2, a_3, x_0, x_1, x_2, x_3\}$ and computes:
$$u=x_0^{-1}a_1x_1, \quad v=x_1^{-1}a_2x_2, \quad w=x_2^{-1}a_3 x_3$$
The private key is $(a_1, a_2, a_3, x_0, x_1, x_2, x_3)$ and the public key is $(u, v, w)$.

(5) Bob chooses $\{b_1, b_2, b_3, y_0, y_1, y_2, y_3\}$ and computes:
$$p=y_0^{-1}b_1y_1, \quad q=y_1^{-1}b_2y_2, \quad r=y_2^{-1}b_3 y_3$$
The private key is $(b_1, b_2, b_3, y_0, y_1, y_2, y_3)$ and the public key is $(p, q, r)$.

(6) Both exchange public keys and compute the following tokens
(Alice) $t_A = a_1 p a_2 q a_3 r = a_1 y_0^{-1} b_1 a_2 b_2 a_3 b_3 y_3 = y_0^{-1} z y_3$
(Bob) $t_B = u b_1 v b_2 w b_3 = x_0^{-1} a_1 b_1 a_2 b_2 a_3 b_3 x_3 = x_0^{-1} z x_3$

where $z = a_1 b_1 a_2 b_2 a_3 b_3$

(7) Both exchange the tokens and compute the common key:
(Alice) $K_A = x_0 \, t_B \, x_3^{-1}$
(Bob) $K_B = y_0 \, t_A \, y_3^{-1}$

Neither Alice, Bob or Mallory (an adversary) could obtain private key parts from the public ones unless they solve triple decomposition problem.

(8) The computed session key could be used to cipher any matrix belonging to the Modular d-dimensional Matrix monoid $(M_d)$, using the Blind Conjugacy Search Problem (BCSP) as OWTF [6]:

***Computational BCSP:***
***Given G, a non-abelian group with solvable word problem and given any element $a \in G$, and an unknown element $b \in G$, find at least one element $x \in G$ such that $a = x^{-1} b \, x$.***

The described cryptosystem goes as:

$$msg \in M_d$$
$$cif = K_B^{-1} msg \, K_B$$
$$dec = K_A \, cif \, K_A^{-1} =$$
$$= K_A (K_B^{-1} msg \, K_B) K_A^{-1} = msg$$

## 4. XTDP SAMPLE PROTOCOL

TABLE I
SYMBOLS AND DEFINITIONS

| |
|---|
| $M_8 = M(8, F_{251})$ –Modular matrix monoid, d=8 and prime=251 |
| $GL(8, F_{251})$ –General Linear Group |
| $\subset$ – strictly included into |
| $\in$ – belongs to |
| $\in_R$ – uniform distribution, randomly selected element in |
| $\forall \neq$ - strictly positive and non-repeating elements in a list. |
| $d_A, d_B, ...$ – diagonal matrices of eigenvalues with $\forall \neq$ property |
| $(\lambda_1 ... \lambda_8)$ – eigenvalues set, each one mentioned independent from others |
| $[a,b]$ – commutator ($=a^{-1}b^{-1}ab$) |
| $I$ – Identity matrix order 8 |
| $Sel$ – selects or reserves for her/him with agreement of the other party. |
| $\Rightarrow$ send publicly to the other entity |

TABLE II
PUBLIC SETUP STEPS

| | |
|---|---|
| Public parameters (any entity defines) | Subgroups $\{A_1,A_2,A_3,X_1,X_2\} \subset M_8$ <br><br> Subgroups $\{A_1,A_2,A_3,X_0,X_1,X_2,X_3\} \subset GL(8,\mathbb{Z}_{251})$ <br> Subgroups $\{B_1,B_2,B_3,Y_0,Y_1,Y_2,Y_3\} \subset GL(8,\mathbb{Z}_{251})$ <br> Eigenvectors O assigned to $A_1$, $Y_0$ ; $[A_1,Y_0]=I$ <br> Eigenvectors P assigned to $A_2$, $Y_1$ ; $[A_2,Y_1]=I$ <br> Eigenvectors Q assigned to $A_3$, $Y_2$ ; $[A_3,Y_2]=I$ <br> Eigenvectors R assigned to $B_1$, $X_1$ ; $[B_1,X_1]=I$ <br> Eigenvectors S assigned to $B_2$, $X_2$ ; $[B_2,X_2]=I$ <br> Eigenvectors T assigned to $B_3$, $X_3$ ; $[B_3,X_3]=I$ <br><br> Eigenvectors public matrices <br> $\{O,P,Q,R,S,T\} \in_R GL(8,\mathbb{Z}_{251}) \Rightarrow$ |

TABLE III
ADDITIONAL ENCRYPTION STEP (BCSP PROTECTED)

| | PRIVATE PROCEDURES | |
|---|---|---|
| | ALICE | BOB |
| Random private elements | Sel $\{A_1,A_2,A_3,X_0,X_1,X_2,X_3\}$<br>$\forall \neq \lambda_1 \ldots \lambda_8 \in_R \mathbb{Z}_{251}^*$<br>$d_{A1} = (\lambda_1 \ldots \lambda_8)$<br>$d_{A2} = (\lambda_1 \ldots \lambda_8)$<br>$d_{A3} = (\lambda_1 \ldots \lambda_8)$<br>$d_{X1} = (\lambda_1 \ldots \lambda_8)$<br>$d_{X2} = (\lambda_1 \ldots \lambda_8)$<br>$d_{X3} = (\lambda_1 \ldots \lambda_8)$<br>$a_1 = O^{-1} d_{A1} O$<br>$a_2 = P^{-1} d_{A2} P$<br>$a_3 = Q^{-1} d_{A3} Q$<br>$x_0 \in_R GL(8, \mathbb{Z}_{251}^*)$<br>$x_1 = R^{-1} d_{X1} R$<br>$x_2 = S^{-1} d_{X2} S$<br>$x_3 = T^{-1} d_{X3} T$ | Sel $\{B_1,B_2,B_3,Y_0,Y_1,Y_2,Y_3\}$<br>$\forall \neq \lambda_1 \ldots \lambda_8 \in_R \mathbb{Z}_{251}^*$<br>$d_{B1} = (\lambda_1 \ldots \lambda_8)$<br>$d_{B2} = (\lambda_1 \ldots \lambda_8)$<br>$d_{B3} = (\lambda_1 \ldots \lambda_8)$<br>$d_{Y0} = (\lambda_1 \ldots \lambda_8)$<br>$d_{Y1} = (\lambda_1 \ldots \lambda_8)$<br>$d_{Y2} = (\lambda_1 \ldots \lambda_8)$<br>$b_1 = R^{-1} d_{B1} R$<br>$b_2 = S^{-1} d_{B2} S$<br>$b_3 = T^{-1} d_{B3} T$<br>$y_0 = O^{-1} d_{Y0} O$<br>$y_1 = P^{-1} d_{Y1} P$<br>$y_2 = Q^{-1} d_{Y2} Q$<br>$y_3 \in_R GL(8, \mathbb{Z}_{251}^*)$ |

TABLE IV
PRIVATE KEYS

| | ALICE | BOB |
|---|---|---|
| Session set of Private keys | $a_1, a_2, a_3, x_0, x_1, x_2, x_3$ | $b_1, b_2, b_3, y_0, y_1, y_2, y_3$ |

TABLE V
PUBLIC KEYS

| | ALICE | BOB |
|---|---|---|
| Public keys: 1st exchange | $u = x_0^{-1} a_1 x_1$<br>$v = x_1^{-1} a_2 x_2$<br>$w = x_2^{-1} a_3 x_3$<br>$(u, v, w) \Rightarrow$ | $p = y_0^{-1} b_1 y_1$<br>$q = y_1^{-1} b_2 y_2$<br>$r = y_2^{-1} b_3 y_3$<br>$(p, q, r) \Rightarrow$ |

TABLE VI
TOKEN GENERATION
*commuting pairs are between brackets*

| | ALICE | BOB |
|---|---|---|
| Token: 2nd exchange | $t_A = a_1 p \, a_2 q \, a_3 r =$<br>$[a_1 y_0^{-1}] b_1 [y_1 a_2] y_1^{-1} b_2 \cdot$<br>$\cdot [y_2 a_3] y_2^{-1} b_3 y_3 =$<br>$y_0^{-1} z \, y_3 \Rightarrow$<br>where<br>$z = a_1 b_1 a_2 b_2 a_3 b_3$ | $t_B = u b_1 v b_2 w b_3 =$<br>$x_0^{-1} a_1 x_1 [b_1 x_1^{-1}] a_2 x_2 \cdot$<br>$\cdot [b_2 x_2^{-1}] a_3 [x_3 b_3] =$<br>$x_0^{-1} z \, x_3 \Rightarrow$<br>where<br>$z = a_1 b_1 a_2 b_2 a_3 b_3$ |

TABLE VII
COMMON SESSION KEY

| | ALICE | BOB |
|---|---|---|
| Common session key | $K_{ALICE} = x_0 \, t_B \, x_3^{-1}$ | $K_{BOB} = y_0 \, t_A \, y_3^{-1}$ |
| | $K_{ALICE} = x_0 \, t_B \, x_3^{-1} =$<br>$= x_0 x_0^{-1} z \, x_3 x_3^{-1} = a_1 b_1 a_2 b_2 a_3 b_3$<br>$K_{BOB} = y_0 \, t_A \, y_3^{-1} =$<br>$= y_0 y_0^{-1} z \, y_3 y_3^{-1} = a_1 b_1 a_2 b_2 a_3 b_3$<br>$\boldsymbol{K_{ALICE} = K_{BOB} \equiv K}$ | |

TABLE VIII
ADDITIONAL ENCRYPTION STEP (BCSP PROTECTED)

| | ALICE | BOB |
|---|---|---|
| BOB ciphers a message to ALICE | | $msg \in M_8$<br>$cif = K_{BOB}^{-1} msg \, K_{BOB}$<br>$cif \Rightarrow$ |

TABLE IX
ALICE DECRYPTION STEP

| | ALICE | BOB |
|---|---|---|
| ALICE recovers the message | $msg =$<br>$= K_{ALICE} \, cif \, K_{ALICE}^{-1}$ | |
| | $msg = K_{ALICE} \, cif \, K_{ALICE}^{-1} =$<br>$= K_{ALICE}(K_{BOB}^{-1} msg \, K_{BOB})K_{ALICE}^{-1} = msg$ | |

It is easy to apply the same framework to other asymmetric protocols. For example, defining power-sets of matrices, a straightforward ElGamal solution is at hand. Also, extending the GL() to a polynomial ring, a Maze et al. protocol could be implemented. Changing of purpose, a ZKP authentication protocol, a Baumslag et al. KEM (key encapsulation mechanism) or a Digital Signature is almost trivial to design. When powers of matrices are used new kind of weakness could appear, the multiplicative order of the elements should be sufficiently high to foil brute-force attacks. One could use companion matrices of primitive polynomials as generators of high order subgroups. Comparing different canonical approaches, the BCSP application of this framework offer a reasonable compromise solution between cryptographic security and fast computation.

Supposing no other weakness at hand, full cracking a private key depends on four *d*-dimensional diagonal eigenvalues matrices, so a brute force search of the commutative $P_8$ subgroups of $M_8$ involves the cardinal

$$|P_8|^4 = 249^{32} = 4.77 \times 10^{76} \approx 2^{255}$$

For a real-life application, we suggest to use at least $P_8$ or perhaps expanding the commutative subgroup to $P_{16}$, who implies a *510*-bit level classical security. The corresponding quantum security level is the respective square roots of the key space cardinals [7]. The obvious drawback is the corresponding increase in space, each matrix in $\{P_8, P_{16}\}$ occupies respectively $\{512, 2048\}$ bits. A corresponding computational session time should be expected.

Against quantum-based attacks, the dimensional increase foils Grover like attacks and no multiplicative (or additive) cyclic order finding adaptation of Shor's algorithm is known and accessible against the present protocol.

Further arguments of the framework security are exposed in our previous work [6].

We designed XTDP to protect the framework against Tsaban's ASA attack [8]. No formal proof is provided, but the presence of full quadratic public terms let us assume it.

As a final consequence, we infer that our proposed framework could be gotten immune against brute-force attacks, linear representation attacks, length-based attacks and currently known quantum attacks.

5. SEMANTIC SECURITY (IND-CCA2)

The key point to assure semantic security is the indistinguishability of encrypted information from random one of same length [9].

The presented framework is easily translatable to other asymmetric protocols. For this reason, the following security analysis is not limited to the present example and can be extended in new contexts. To proceed with this analysis, two conjectures are exposed.

### *DEFINITION (D1): interactive challenge-response game by a verifier against an active adversary.*

In this three-phase protocol, two entities, an adversary and a verifier (or challenger) are involved. The verifier has a secret key that he tries to hide from the adversary and allows the adversary to pose questions to him that answers truthfully like an Oracle. In a first phase, the adversary can raise all the questions that he wants to try to obtain information about the secret key. In a second phase, the verifier presents the secret key (k) to the adversary next to another of equal length and format (* k) randomly generated. Even during the second phase, the adversary may continue to consult the verifier, except for questions linked to the disclosure of the secret itself. In the third phase, the adversary has a polynomial time stochastic algorithm and must distinguish whether the secret is k or * k, with probability negligibly greater than ½. If the opponent achieves the distinction with that probability, he wins the game and loses it in the opposite case.

### *CONJECTURE (C1): Indistinguishability of product transformed random matrices.*

The elements of $GL(d, F_p)$ are uniformly random integers of prime modulus and d-dimension. It is a known fact that sum or multiplication between random field integers, does not introduce statistical bias into results. Therefore, linear transformed matrices are statistically distributed as any random generated ones. The consequence is that in an interactive challenge-response protocol (Definition D1), an adversary does not achieve the distinction raised with the required probability.

### *CONJECTURE (C2): the present framework adheres to semantic security under IND-CCA2.*

The TDP one-way trapdoor function with which the private key is protected is, as previous exposed, not weakened by attacks of probabilistically polynomial time, whether classical or quantum that are in the public domain until today, forcing the potential attacker to perform a systematic exploration of the private key space (the diagonal matrices). Under this assumption and considering the indistinguishability of randomly generated matrices and enciphered ones, it is reasonable to assign to the framework a security mark equivalent to IND-CCA2 (semantic security under IND-CCA2) [9].

### 6. CONCLUSION.

We developed a fortified and supposedly ASA [8] resistant version of our previous TDP framework. Evidence is provided to adhere to IND-CCA2 semantic security.

### 7. ACKNOWLEDGMENT.

We are indebted to Boaz Tsaban and Yuval Beker for letting us take contact with their work and the following discussions.

### 8. REFERENCES.